\documentclass[10pt]{article} 
\usepackage{times}
\usepackage{epsfig}
\usepackage{color}
\usepackage{graphicx}
\usepackage{url}
\usepackage{multirow}
\usepackage{amsmath, amsthm, amssymb} 
\usepackage{epsfig} 
\usepackage{setspace}

\newcommand{\hide}[1]{}

\newcommand{\calS}{{\mathcal S}}

\newtheorem{theorem}{Theorem} 
\newtheorem{observation}{Observation}
 
\newtheorem{lemma}{Lemma} 
\newtheorem{conjecture}{Conjecture}

\newcommand{\comment}[1]{} 

\begin{document}

\title{Improved Approximation Algorithms for Segment Minimization in Intensity Modulated Radiation 
Therapy\thanks{Research partially supported by NSERC.}}

\author{\hspace{1.5cm}Therese Biedl\thanks{David R. Cheriton School of Computer Science, 
University of Waterloo, ON, Canada,\{biedl, 
m22young\}@uwaterloo.ca}
\and Stephane Durocher\thanks{Department of Computer Science, 
University of Manitoba, MB, Canada, durocher@cs.umanitoba.ca}
\and Holger H.~Hoos\thanks{Department of Computer Science, University of British Columbia, 
BC, Canada, hoos@cs.ubc.ca}~~~~~~~~~~~~~~~~ \and 
Shuang Luan\thanks{Department of Computer Science, 
University of New Mexico, NM, USA,\{sluan, saia\}@cs.unm.edu} \and Jared Saia$^{\P}$ \and Maxwell 
Young$^{\dag}$}

\maketitle

\begin{abstract}
The segment minimization problem consists of finding the smallest
set of integer matrices that sum to a given intensity matrix, such 
that each summand has only one non-zero value, and the non-zeroes 
in each row are consecutive. This has direct applications in 
intensity-modulated radiation therapy, an effective form of cancer 
treatment. We develop three approximation algorithms for matrices with arbitrarily many
rows. Our first two algorithms improve the approximation factor from the previous
best of $1+\log_2 h $ to (roughly)
$3/2 \cdot (1+\log_3 h)$ and $11/6\cdot(1+\log_4{h})$, respectively,
where $h$ is the largest entry in the intensity matrix. We illustrate the limitations of
the specific approach used to obtain these two algorithms
by proving a lower bound of $\frac{(2b-2)}{b}\cdot\log_b{h} + 
\frac{1}{b}$ on the approximation guarantee. Our third algorithm 
improves the approximation factor from 
$2 \cdot (\log D+1)$ to $24/13 \cdot (\log D+1)$, where $D$ is 
(roughly) the largest difference between consecutive elements of a 
row of the intensity matrix.  Finally, experimentation with these algorithms 
shows that they perform well with respect to the optimum and 
outperform other approximation algorithms on 77\% of the 122 test 
cases we consider, which include both real world and synthetic 
data.
\end{abstract}

\vspace*{-2mm}
\section{Introduction}

Intensity-modulated radiation therapy (IMRT) is an effective form 
of cancer treatment in which the region to be treated is 
discretized into a grid and a treatment plan specifies the 
amount of radiation to be delivered to the area of body surface 
corresponding to each grid cell. A device called a multileaf 
collimator (MLC) is used to administer the treatment plan in a 
series of steps. In each step, two banks of metal leaves in the 
MLC are positioned to cover certain portions of the body surface, 
while leaving others exposed, and the latter are then subjected to 
a specific amount of radiation.

A treatment plan can be represented as an $m \times n$ 
\emph{intensity matrix} $T$ of non-negative integer values, whose 
entries represent the amount of radiation to be delivered to the 
corresponding grid cells. The leaves of the MLC can be seen as 
partially covering rows of $T$; for each row $i$ of $T$ there are 
two leaves, one of which may slide inwards from the left to cover 
the elements in columns $1..l$ of that row, while the other may 
slide inwards from the right to cover the elements in columns 
$r..n$. After each step of the treatment, the amount of radiation 
applied in that step (this can differ per step) is subtracted from 
each entry of $T$ that has not been covered. The treatment is 
completed when all entries of $T$ have reached $0$.

Setting leaf positions in each 
step of the treatment plan requires time. Minimizing the number of 
steps reduces treatment time and can result in 
increased patient throughput, reduced machine wear 
and tear, and overall reduced cost of 
the procedure. Minimizing the number of steps 
for a given treatment plan is the objective of this paper.

Formally, a {\it segment} is a matrix $S$ such that non-zeroes in 
each row of $S$ are consecutive, and all non-zero entries of $S$ 
are the same integer, which we call the {\em segment-value}. A 
\emph{segmentation} of $T$ is a set of segment matrices that sum 
to $T$, and we call the cardinality of such a set the 
\emph{size} of that segmentation. The {\em segmentation problem} 
is, given an intensity matrix $T$, to find a minimum-size 
segmentation of $T$. We will often consider the special case of a 
matrix $T$ with one row, which we call the {\em single-row 
segmentation problem} as opposed to the {\em full-matrix 
segmentation problem.}

The segmentation problem is known to be NP-complete in the strong 
sense, even for a single row 
\cite{baatar:new,baatar:decomposition,chen:generalized}, as well 
as APX-complete~\cite{B}. A number of heuristics are 
known~\cite{baatar:thesis,baatar:decomposition,cotrutz:segment,engel:new,siochi:minimizing,xia:multileaf}. Approaches
for obtaining optimal (exact) solutions also
exist~\cite{baatar:minimum,brand:sum,kalinowski:complexity,wakea:mixed};
of course, these approaches do not necessarily terminate in polynomial time in the
size of the input. Bansal~{\it  
et al.}~\cite{B} provide a $24/13$-approximation algorithm for the 
single-row problem and give some better approximations for more 
constrained versions. Collins~{\it et 
al.}~\cite{collins:nonnegative} show that the single {\it column} 
version of the problem is NP-complete and provides some 
non-trivial lower bounds given certain constraints. Work by 
Luan {\it et al.}~\cite{luan:approximation} gives two 
approximation algorithms for the full $m\times{}n$ problem where the
approximation factor
depends on other parameters of the problem, e.g.~the 
largest entry $h$ in the target matrix.
They do not consider the performance of their algorithms 
in practice. More recent work
by~\cite{kalinowski:complexity} has shown that the 
$m\times{}n$ case can be solved optimally with time complexity
$O(m\cdot{}n^{2h+2})$; this approach is shown to computationally
intensive even for small $h$ in practice. 
   
\vspace{-3mm}

\subsection*{Our Contributions}

Luan et al.~\cite{luan:approximation} used two properties to obtain 
approximation algorithms. First, the segmentation problem is straightforward
when $h=1$ (0/1-matrices). Second, segmentations for the single-row problem 
with small segment-values can be used to obtain good segmentations 
for the full-matrix problem. By exploiting these two properties, Luan
{\it et al.} obtained two 
algorithms with respective approximation factors of 
$1+\log_2 h$ and $2(1+\log_2 D)$ where $h$ is the largest value in $T$, and 
$D$ is roughly the largest difference between consecutive elements 
in a row of $T$.\footnote{Throughout, we use $\log_{b}x$ to mean $\lceil{}\log_{b}x\rceil$.}

In this paper, we extend the ideas of Luan~{\it et al}. In particular, we
prove that the segmentation problem can be approximated when $h=2$
and $h=3$; this is far less straightforward than the case $h=1$.  
This yields two fast algorithms for the full-matrix segmentation problem with 
approximation factors (roughly) $\frac{3}{2} \cdot (1+\log_3{h})$ and
$\frac{11}{6}\cdot{}(1+\log_4{h})$, respectively,  
both of which are less than $1+\log h$. While we 
show that the general two-stage approach of Luan {\it et
al}.~\cite{luan:approximation} can be extended to provide superior
approximation algorithms, we also prove a limitation of this
approach.  

We also provide a new approximation algorithm with approximation factor 
(roughly) $\alpha\log D$, where $\alpha$ is the best approximation 
factor for the single-row problem. The current best known $\alpha$ 
is $\alpha=24/13$ \cite{B}; any improved approximation result for 
the single-row problem would directly lead to an improved 
approximation result for the full problem. This second 
approximation algorithm expands on the second approximation 
algorithm by Luan et al.; they used one specific 2-approximation 
algorithm for the single-row problem, whereas we show that in fact 
any $\alpha$-approximation algorithm can be used.

Finally, we give an empirical evaluation of known approximation 
algorithms for the full $m\times{}n$ segmentation problem, using 
both synthetic and real-world clinical data. Our experiments 
demonstrate that the constant factor improvements made by our 
algorithms yield significant performance gains in practice. 
Therefore, in both the $O(\log{h})$ and $O(\log{D})$ scenarios, 
our new algorithms improve on previous approximation algorithms 
theoretically and experimentally.

\section{Improved Approximation Algorithms}\label{section:logh}

A vital insight for our approximation algorithm is the concept
of a {\em marker} (\cite{luan:approximation}; this was 
called {\em tick} in \cite{B}.)  
A {\em marker} in row $i$ of the target matrix $T$ is an index 
where the entry of $T$ changes while going along the row.  Formally,
it is an index
$j$ for which $T[i,j-1]\neq T[i,j]$, or $j=1$ and $T[i,1]\neq 0$,
or $j=n+1$ and $T[i,n]\neq 0$.  

Let $\rho^{i}$ denote the number of markers in row
$i$ of $T$, and define $\rho = \max\limits_{\mbox{\tiny{All
      rows $i$ in $T$}}}\{\rho^i\}$, i.e.  the number
of markers in the row of $T$ which has the most markers over all rows.
 We begin by restating the following observation noted by Luan~{\it
et al.} that we will later find useful.

\begin{observation}\label{observation:OPT}
(Luan~{\it et al.}~\cite{luan:approximation})~Let $OPT$ be the size of a minimal segmentation of an intensity matrix
$T$. Then $\rho \leq{} 2\cdot{}OPT$.
\end{observation}

The first approximation algorithm given by Luan et 
al.~\cite{luan:approximation} works as follows. Split the given 
intensity matrix $T$ into matrices $P_0,\dots,P_k$ such that 
$T=\sum_{\ell=0}^k 2^\ell \cdot P_\ell$ (by taking the bits of the 
base-2 representation of entries of $T$) where $k=\log_2{h}$ and each
$P_{\ell}$ is a $0/1$-matrix. A segmentation for $T$ 
can then be obtained by taking segmentations of each $P_\ell$, 
multiplying their values by $2^\ell$, and taking their union. 
Since each $P_\ell$ is a 0/1-matrix, an optimal segmentation of it 
can be found easily, and an approximation bound of $1+\log h$ 
can be shown.

We use a similar approach, but change the base $b$, writing 
$T=\sum_{\ell=0}^k b^\ell \cdot P_\ell$ for some integer $b\geq 
3$. 
This raises nontrivial 
question: How can we solve the segmentation problem in a matrix 
that has values in $\{0,1,\dots,b-1\}$?  And is the resulting 
segmentation a good approximation of the optimal segmentation?

Assume that we have {\em $\alpha$-approximate segmentations} for each $P_\ell$,
i.e., for each $\ell$ we have a segmentation $\calS_\ell$ of $P_\ell$ that
is within a factor $\alpha$ of the optimum for $P_\ell$, for 
some $\alpha\geq 1$.
We {\em combine} these segmentations as follows:  For each segment $S$ of 
$\calS_\ell$, add $b^\ell\cdot S$
to $\calS$.  One easily verifies that $\calS$ is a segmentation of $T$.
But it is not obvious that this is a good approximation of the optimum
segmentation of $T$.  One might think that it is an 
$\alpha(\log_b(h)+1)$-approximation of the optimal segmentation of $T$,
but this is {\em not} true in general; see also
Section~\ref{se:higher_base}.  

It is also not clear how to find a segmentation of $P_\ell$ that is
good.  As mentioned earlier, the optimal segmentation can be found
in polynomial time if $b$ is a constant \cite{kalinowski:complexity},
but the running time is not practical, and it is not clear whether
it yields a good approximation.  Our main contribution is that an
approximation
guarantee can be established for $b=3,4$.   Moreover, it suffices to
use a segmentation of $P_\ell$ that is not necessarily optimal, but
can be found in linear time.

More specifically,
we show how to find a segmention of one row of $P_\ell$
that can be bound in size depending
on the number of markers $\rho$.  Moreover, the segmentations of
each row can be combined easily into one segmentation of $P_\ell$,
and the segmentations of all the $P_\ell$'s can be combined into
a segmentation of $T$, while carrying the bound in terms of
$\rho$ along.  By Observation~\ref{observation:OPT}, this will
allow us to bound the size of resulting segmentation relative to
the optimum.

We briefly give here the simple algorithm {\sc GreedyRowPacking}
that we use to combine segmentations of rows of a target-matrix $P_\ell$
(with values in $1,\dots,b-1$)
into a segmentation of the whole matrix $P_\ell$.
Check for each value $v\in \{1,\dots,b-1\}$ whether
any segment in any row has this value. If there is one, then remove a
segment of value $v$ from each row that has one. Combine all these
segments into one segment-matrix (also with value $v$), and add it to
$\calS$.   Continue until all segments in all rows have been used
in a segment-matrix.  Clearly if each row has at least $n_i$ {\em $i$-segments}
(i.e., segments with value $i$), then {\sc GreedyRowPacking} gives
a segmentation of $P_\ell$ with at most $n_i$ $i$-segments (and
$n_1+\dots+n_{b-1}$ segments in total.)

\subsection{Basis $b=3$}

We now explain in detail the approach when the target-matrix has been
split by base $b=3$.  Thus, we are now interested in obtaining a
segmentation of an intensity matrix $P_\ell$ 
that has all entries in $\{0,1,2\}$; we
call this a {\em $0/1/2$-matrix}.  Recall that
$\rho^i$ is the number of markers in the $i$th row of the target matrix $T$.
We use $\rho_\ell^i$ to denote he number of markers in the $i$th row
of $P_\ell$.

\begin{lemma}
\label{lem:markers}
There exists a segmentation of row $i$ of a $0/1/2$-matrix $P_\ell$ such
that the number of 1-segments is at most $\frac{1}{2}\cdot \rho_\ell^i$,
and the number of 2-segments is at most $\frac{1}{4} \cdot \rho_\ell^i+\frac{1}{2}$.
\end{lemma}
\begin{proof}
We prove this by induction on $\rho_\ell^i$.  The base case will be that none
of the cases for the induction can be applied, and hence will be treated
last.  For the induction,
we prove this by repeatedly identifying a subsequence of the row 
for which we can add a few segments and remove many markers, where 
``remove'' means that if we subtracted the segments from the 
target row, we would have fewer markers.   To identify subsequences 
of the row, we use regular expression notation.  The bound then
follows by induction.

We will give this in detail only for the first of the cases in the induction
step, and only briefly sketch the others:
\begin{enumerate}
\item Assume that the row contains a subsequence of the form $12^+1$.
	Let $s$ be a 1-segment that covers exactly the subsequence of $2$s,
	and consider $P'=P-s$.  Then $P'$ has two fewer markers in the $i$th
	row (at the endpoints of $s$), and so by induction the $i$th row
	can be segmented using at most 
	$\frac{1}{2}\cdot (\rho_\ell^i-2)$ 1-segments,
	and $\frac{1}{4} \cdot (\rho_\ell^i-1)+\frac{1}{2}$ 2-segments.
	Adding the 1-segment $s$ to this segmentation yields the desired
	result.
\item If there exists a subsequence of the form $01^+0$,
    then similarly apply a 1-segment at the subsequence of $1$s.  This removes
    2 markers, and adds one 1-segment, and no 2-segment to the inductively
	obtained segmentation.
\item If there exists a subsequence of the form $02^+1^+2^+0$,
    then similarly apply a 2-segment at the first subsequence of $2$s, then two
    1-segments to remove the remaining $1^+2^+$.  This removes
    4 markers, and adds two 1-segments, and one 2-segment to the inductively
	obtained segmentation.
\item If there exist two subsequences of the form 
$02^+1^+0$
    or $01^+2^+0$, then similarly apply one 1-segment to one subsequence of $2$s,
    and one 2-segment to the other subsequence of $2$s, then apply
    two 1-segments to the two remaining sequences of $1$s.  This
    removes 6 markers, and adds three 1-segments and one 2-segment to the
	inductively obtained segmentation.
\item If there exist two subsequences of the form $02^+0$,
    then similarly apply one 2-segment to one of them, and two 1-segments to the 
    other.  This
    removes 4 markers, and adds two 1-segments and one 2-segment to the 
	inductively obtained segmentation.
\item If there exists one subsequence of the form $02^+1^+0$
    or $01^+2^+0$, and one subsequence of the form $02^+0$,
    then similarly apply one 2-segment to the subsequence $02^+0$, and two one 
    1-segments to the other subsequence.  This removes 5 markers,
    and adds two 1-segments and one 2-segment to the inductively obtained
	segmentation
\end{enumerate}

Now assume that none of the above cases can be applied (i.e., the base case.)
We argue that in fact at most three markers are left. Let 
$0(1+2)^+0$ be a subsequence that has markers in it. Assume first 
the leftmost non-zero is a 1. Then the subsequence must contain a 
2 somewhere (otherwise we're in case (2)), so it has the form 
$01^+2^+(1+2)^+0$. But after the 2s, no 1 can follow (otherwise 
we're in case (1)), so this subsequence has the form $01^+2^+0$. 
Likewise, if the last non-zero is 1, then the subsequence has the 
form $02^+1^+0$. If the first and last non-zero are 2, then the 
subsequence has the form $02^+0$ (otherwise we're in case (1) or 
(3)).

If we had two subsequences $0(1+2)^+0$, then each would have the 
form $01^+2^+0$ or $02^+1^+0$ or $02^+0$, and we would be in case 
(4), (5) or (6). So there is only one of them, and it has at most 
three markers. We can now eliminate either three remaining markers with a 
1-segment and a 2-segment, or two remaining markers with a 
2-segment; either way the bound holds. 
\end{proof}

\begin{figure}[ht]
\hspace*{\fill}
\input{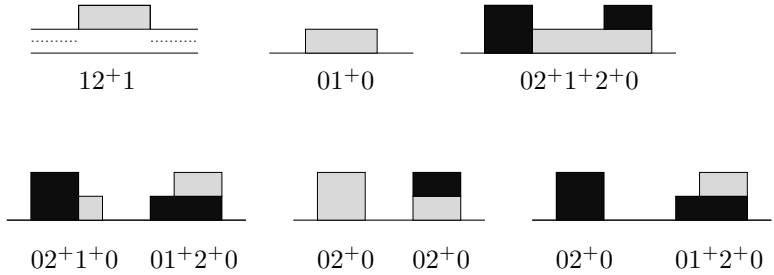}
\hspace*{\fill} \caption{A segmentation where the number of 
segments is bounded by markers. This illustrates cases (1) through (6)
of the proof of Lemma~\ref{lem:markers}.}
\end{figure}

Using the segmentations of each row obtained with Lemma~\ref{lem:markers},
and combining them with algorithm {\sc GreedyRowPacking}, gives a segmentation
$\calS_\ell$
of each 0/1/2-matrix $P_\ell$. 
We now show that combining these segments
gives a provably good
approximation of the optimal segmentation of $T$.

\begin{lemma}
Assume $T=\sum_{\ell=0}^k 3^\ell P_\ell$, where $k=\log_3 h$
and each $P_\ell$ is a 0/1/2-matrix. Combining  the above
segmentations $\calS_0,\allowbreak\dots,\calS^*_k$ for matrices 
$P_0,\dots,P_k$ gives a segmentation $\calS$ for $T$ of size at 
most $\frac{3}{2} \cdot k\cdot OPT+ \frac{1}{2} \cdot k$, where 
$OPT$ is the size of a minimal segmentation of $T$.
\end{lemma}
\begin{proof}
Recall that the segmention of
row $i$ of $P_\ell$ has at most
$\frac{1}{2} \cdot \rho_\ell^i$ 1-segments
and at most
$\frac{1}{4} \cdot \rho_\ell^i+\frac{1}{2}$ 2-segments
(Lemma~\ref{lem:markers}).
Let $\rho_\ell=\max_i \rho_\ell^i$ be the maximum number of markers within
any row of $P_\ell$.  
By algorithm {\sc GreedyPacking} segmentation
$\calS_\ell$ of $P_\ell$ then has at most
$\frac{1}{2} \cdot \rho_\ell$  1-segments
and at most
$\frac{1}{4} \cdot \rho_\ell+\frac{1}{2}$ 2-segments.  
So
$$|\calS_\ell| \leq \frac{3}{4} \cdot \rho_\ell + \frac{1}{2}.$$
Matrix $P_\ell$ can have a marker only if matrix $T$
has a marker in the same location, so $\rho_\ell \leq \rho$
\cite{luan:approximation}.  By Observation~\ref{observation:OPT},
$\rho\leq 2\cdot OPT$.
Putting it
all together, we have
$$|\calS| 
= \sum_{\ell=0}^k |\calS_\ell|
\leq \sum_{\ell=0}^k \left(\frac {3}{4} \cdot \rho_\ell + \frac{1}{2}\right)
\leq \sum_{\ell=0}^k \left(\frac {3}{4}\cdot{}2\cdot OPT + \frac{1}{2}\right)
 = \left(\frac {3}{2}\cdot{}OPT + \frac{1}{2}\right)\cdot{}\left(1+\log_{3}{h}\right)$$
which proves the result.  
\end{proof}

The above result showed the approximation bound for the segmentation 
obtained by packing the segmentations of the rows of Lemma~\ref{lem:markers}
into matrices. For each matrix $P_{\ell}$, this requires $O(m\cdot{}n)$ time; 
therefore, 
the entire algorithm runs in time $O(m\cdot{}n\cdot{}\log{h})$. 

We note here that in the above proof, one could also have used an optimal
segmentation $\calS_\ell^*$ of $P_\ell$ 
instead of the segmentation $\calS_\ell$;
since $|\calS_\ell^*|\leq |\calS_\ell|$, the same approximation bound holds
for the resulting segmentation of $T$.  However, it is doubtful whether
the increased run-time of $O(mn^6)$ to find the optimal segmentation
\cite{kalinowski:complexity} is worth the improvement in quality. 

We can now restate our result as a theorem:

\begin{theorem}
There exists an algorithm running in $O(m\cdot{}n\cdot{}\log{h})$ 
time that for any intensity matrix $T$ with maximum value $h$ 
finds a segmentation $\calS$ of $T$ size at most $\frac{3}{2} 
\cdot (\log_3 h +1) \cdot OPT + \frac{1}{2} \cdot (\log_3 h+1),$ 
where $OPT$ is the size of a minimal segmentation of $T$.
\end{theorem}

\subsection{Basis $b=4$}
\label{sec:log4h}

\begin{figure}
\centering
\includegraphics[width=0.85\linewidth]{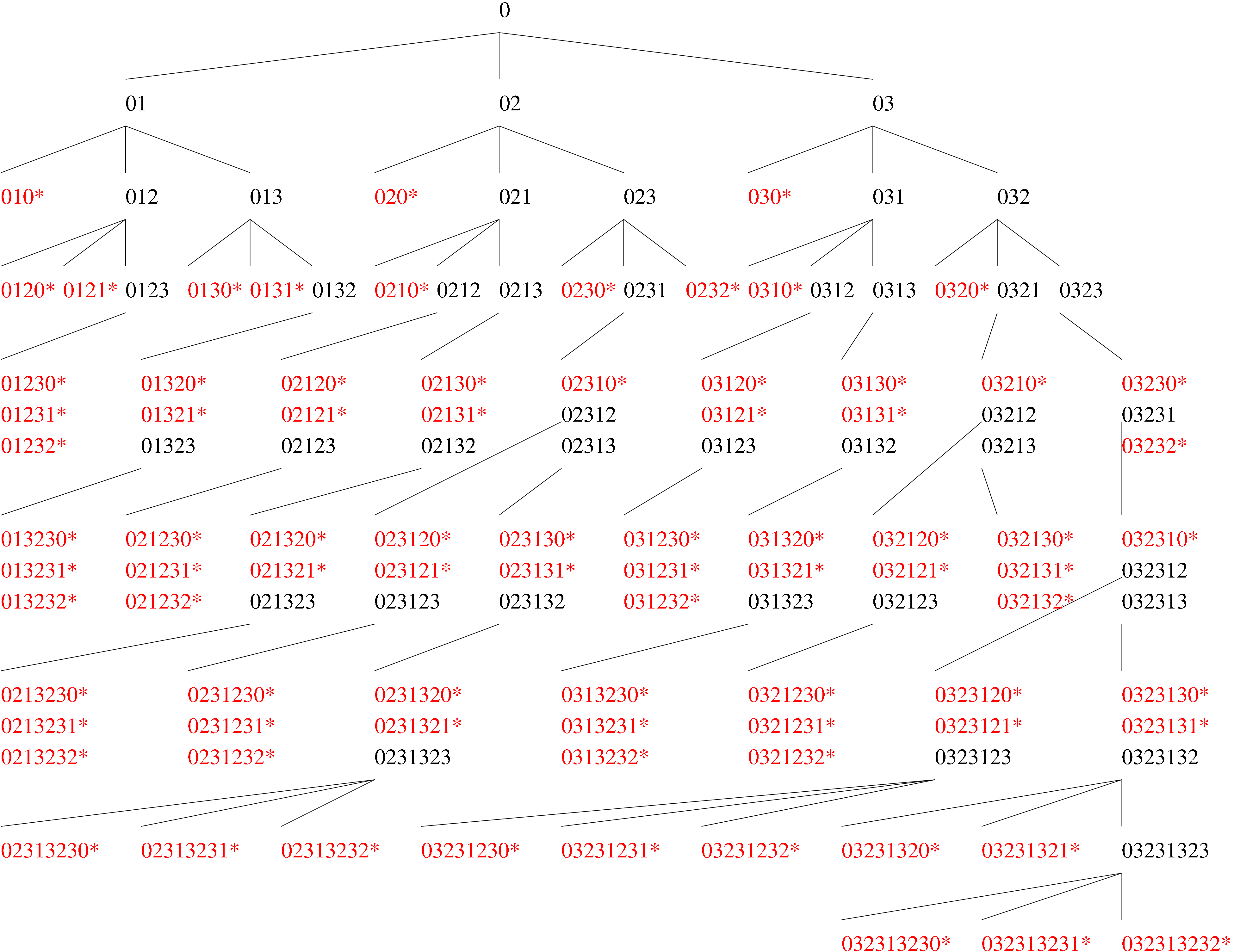}
\caption{Generating all substrings that begin with 0. Substrings that
contain an island are marked with an asterisk and not evaluated further.
Multiple consecutive symbols are
omitted; only the first instance of the symbol is included.}
\label{fig1}
\end{figure}

With an extensive case analysis, we can provide an analogue to
Lemma~\ref{lem:markers} for $b=4$ as well; we provide this analysis
here for completeness. From now on, let $P_\ell$ be a $0/1/2/3$-matrix (a
matrix with entries in $\{0,1,2,3\}$) and as before let 
$\rho_\ell^i$ be the number of markers in row $i$ of $P_\ell$.  
We have the following result:

\begin{lemma}\label{lemma_b_4}
There exists a segmentation of row $i$ of the $0/1/2/3$-matrix $P_\ell$ 
consisting of at most
$\frac{1}{2}\rho_\ell^i+O(1)$ 1-segments, $\frac{1}{4}\rho_\ell^i$ 
2-segments, and $\frac{1}{6}\rho_\ell^i$ 3-segments.  
\end{lemma} 
\begin{proof}
The proof is similar to Lemma~\ref{lem:markers} in structure, and proceeds
by induction on $\rho_\ell^i$.  The base case is that none of the inductive
cases can be applied; we will return to this later.

In the induction step, just as in Lemma~\ref{lem:markers} we search
for subsequences (described by regular expressions), and show how we
can ``remove'' $m_i$ markers from a given subsequence by using at most
$\frac{1}{2}m^i$ 1-segments, $\frac{1}{4}m^i$ 2-segments, and
$\frac{1}{6}m^i$ 3-segments.  As will be apparent, it suffices to
only consider sequences that contain an {\em island},  
where an island is a sequence $s$ that begins and ends with the same
number and has only larger numbers inbetween, i.e., 
there is a unique symbol $\sigma \in \{0,1,2\}$ for which
$s = \sigma^+ ( (\sigma+1)|\cdots|3)^+ \sigma^+$.

We generate the set of possible sequences that begin with $0$ and 
contain at at most one island by considering
the tree whose recursive construction is defined as follows:
\begin{enumerate}
\item
Each node is a sequence over $0(0|1|2|3)^+$.
\item
Set the root to string 0.
\item
If a node contains an island, then that node is a leaf, otherwise it is
an internal node with three children.
\item
If a node $s$ is an interior node with last symbol $\sigma$, then its
children are $s0$, $s1$, $s2$ and $s3$. Since $\sigma\in\{0,1,2,3\}$,
we omit the child whose last two symbols are $\sigma\sigma$, resulting
in only three children. 
\end{enumerate}
The complete tree is illustrated in Figure~\ref{fig1} and each leaf 
contains an island.  In particular, this shows that any subsequence
must contain an island, so it suffices to show how to segment islands.

Table~\ref{tab1} gives a segmentation of each leaf node string (or
multiple copies of that leaf node string)
that respects the bound.  If the island contained in the leaf node
begins with $\sigma> 0$, then the segmentation is the same as for
the island where all values have been decreased by $\sigma$; 
in such cases, Table~\ref{tab1} refers to the matching island.  

We illustrate how
to read this table for case 030 only; all other cases are similar.
Assume there are 6 occurrences of the pattern $03^+0$, which hence
have 12 markers.  Define 6 1-segments, 3 2-segments and 2 3-segments
that together cover these 6 substrings.  Apply induction to the rest
of the row, and add these 11 segments to the resulting segmentation; 
this then gives a segmentation of the $i$th row of $P_\ell$ with
the desired bounds. 

Applying similar arguments to all other cases yields the inductive step.
Since we have covered all possible patterns containing one island, 
the only case remaining
for the base case is that some patterns occurs, but not as often as
demanded in Table~\ref{tab1}.  Since there is a finite number of patterns,
each of which has a finite number of markers, there are hence only $O(1)$
markers left
and clearly this can be covered with $O(1)$ 1-segments.
\end{proof}

\hspace{-0.5cm}\begin{table}
\begin{minipage}{2.8in}
\scriptsize
\begin{tabular}{|r|r|l|l|l|l|l|}
\hline
\multicolumn{2}{|c|}{leaf node} & \multicolumn{5}{|c|}{bounded segmentation} \\
\hline
leaf & island & copies & $\rho$ & 1-seg & 2-seg & 3-seg \\
\hline
\hline
010 & 010 & 1 & 2  & 1 & 0 & 0 \\
020 & 020 & 2 & 4  & 2 & 1 & 0 \\
030 & 030 & 6 & 12 & 6 & 3 & 2 \\
\hline
\hline
0120 & 0120 & 2 & 6 & 3 & 1 & 0 \\
\hline 0121 & 121  & \multicolumn{5}{|l|}{see 010} \\ 
\hline
0130 & 0130 & 2 & 6  & 2 & 1 & 1 \\
\hline 0131 & 131  & \multicolumn{5}{|l|}{see 020} \\
\hline 0210 & 0210 & \multicolumn{5}{|l|}{see 0120} \\ 
\hline
0230 & 0230 & 2 & 6 & 3 & 1 & 1 \\
\hline 0232 & 232  & \multicolumn{5}{|l|}{see 010} \\
\hline 0310 & 0310 & \multicolumn{5}{|l|}{see 0130} \\
\hline 0320 & 0320 & \multicolumn{5}{|l|}{see 0230} \\
\hline
\hline
01230 & 01230 & 1 & 4 & 2 & 1 & 0 \\
\hline 01231 & 1231 & \multicolumn{5}{|l|}{see 0120} \\
\hline 01232 & 232 & \multicolumn{5}{|l|}{see 010} \\
\hline
01320 & 01320 & 1 & 4 & 2 & 1 & 0 \\
\hline 01321 & 1321 & \multicolumn{5}{|l|}{see 0210} \\
\hline
02120 & 02120 & 1 & 4 & 2 & 1 & 0 \\
\hline 02121 & 121 & \multicolumn{5}{|l|}{see 010} \\
\hline
02130 & 02130 & 2 & 8 & 3 & 2 & 1 \\
\hline 02131 & 131 & \multicolumn{5}{|l|}{see 020} \\
\hline 02310 & 02310 & \multicolumn{5}{|l|}{see 01320} \\
\hline 03120 & 03120 & \multicolumn{5}{|l|}{see 02130} \\
\hline 03121 & 121 & \multicolumn{5}{|l|}{see 010} \\
\hline
03130 & 03130 & 2 & 8 & 4 & 2 & 1 \\
\hline 03131 & 131 & \multicolumn{5}{|l|}{see 020} \\
\hline 03210 & 03210 & \multicolumn{5}{|l|}{see 01230} \\
\hline
03230 & 03230 & 1 & 4 & 2 & 1 & 0 \\
\hline 03232 & 232 & \multicolumn{5}{|l|}{see 010} \\
\hline
\hline
013230 & 013230 & 2 & 10 & 3 & 2 & 1 \\
\hline 013231 & 13231 & \multicolumn{5}{|l|}{see 02120} \\
\hline 013232 & 232 & \multicolumn{5}{|l|}{see 010} \\
\hline
021230 & 021230 & 2 & 10 & 5 & 2 & 1 \\
\hline 021231 & 1231 & \multicolumn{5}{|l|}{see 0120} \\
\hline 021232 & 232 & \multicolumn{5}{|l|}{see 010} \\
\hline
021320 & 021320 & 1 & 5 & 2 & 1 & 0 \\
\hline
\end{tabular}
\end{minipage}
\begin{minipage}{2.8in}
\scriptsize
\begin{tabular}{|r|r|l|l|l|l|l|}
\hline
\multicolumn{2}{|c|}{leaf node} & \multicolumn{5}{|c|}{bounded segmentation} \\
\hline
leaf & island & copies & $\rho$ & 1-seg & 2-seg & 3-seg \\
\hline
\hline 021321 & 1321 & \multicolumn{5}{|l|}{see 0210} \\
\hline 023120 & 023120 & \multicolumn{5}{|l|}{see 021320} \\
\hline 023121 & 121 & \multicolumn{5}{|l|}{see 010} \\
\hline 023130 & 023130 & 2 & 10 & 4 & 2 & 1 \\
\hline 032131 & 131 & \multicolumn{5}{|l|}{see 020} \\
\hline 032132 & 2132 & \multicolumn{5}{|l|}{see 0120} \\
\hline 032310 & 032310 & \multicolumn{5}{|l|}{see 013230} \\
\hline
\hline
0213230 & 0213230 & 1 & 6 & 3 & 1 & 1 \\
\hline 0213231 & 13231 & \multicolumn{5}{|l|}{see 02120} \\
\hline 0213232 & 232 & \multicolumn{5}{|l|}{see 010} \\
\hline
0231230 & 0231230 & 1 & 6 & 2 & 1 & 1 \\
\hline 0231231 & 1231 & \multicolumn{5}{|l|}{see 0120} \\
\hline 0231232 & 232 & \multicolumn{5}{|l|}{see 010} \\
\hline
0231320 & 0231320 & 1 & 6 & 3 & 1 & 1 \\
\hline 0231321 & 1321 & \multicolumn{5}{|l|}{see 0210} \\
\hline
0313230 & 0313230 & 1 & 6 & 2 & 1 & 1 \\
\hline 0313231 & 13231 & \multicolumn{5}{|l|}{see 02120} \\
\hline 0313232 & 232 & \multicolumn{5}{|l|}{see 010} \\
\hline
0321230 & 0321230 & 1 & 6 & 3 & 1 & 1 \\
\hline 0321231 & 1231 & \multicolumn{5}{|l|}{see 0120} \\
\hline 0321232 & 232 & \multicolumn{5}{|l|}{see 010} \\
\hline 0323120 & 0323120 & \multicolumn{5}{|l|}{see 0213230} \\
\hline 0323121 & 121 & \multicolumn{5}{|l|}{see 010} \\
\hline 0323130 & 0323130 & \multicolumn{5}{|l|}{see 0313230} \\
\hline 0323131 & 131 & \multicolumn{5}{|l|}{see 020} \\
\hline
\hline
02313230 & 02313230 & 2 & 14 & 6 & 3 & 2 \\
\hline 02313231 & 13231 & \multicolumn{5}{|l|}{see 02120} \\
\hline 02313232 & 232 & \multicolumn{5}{|l|}{see 010} \\
\hline
03231230 & 03231230 & 2 & 14 & 7 & 3 & 2 \\
\hline 03231231 & 1231 & \multicolumn{5}{|l|}{see 0120} \\
\hline 03231232 & 232 & \multicolumn{5}{|l|}{see 010} \\
\hline 03231320 & 03231320 & \multicolumn{5}{|l|}{see 02313230} \\
\hline 03231321 & 1321 & \multicolumn{5}{|l|}{see 0210} \\
\hline
\hline
032313230 & 03231323 & 1 & 8 & 2 & 2 & 1 \\
\hline 032313231 & 13231 & \multicolumn{5}{|l|}{see 02120} \\
\hline 032313232 & 232 & \multicolumn{5}{|l|}{see 010} \\
\hline
\end{tabular}
\end{minipage}
\caption{Segmentations for each leaf in Figure~\ref{fig1}. Horizontal rules
separate leaves from different levels in the tree.}
\label{tab1}
\end{table}

\noindent{}We now have the following theorem:

\begin{theorem}\label{theorem_b_4}
There exists an algorithm running in $O(m\cdot{}n\cdot{}\log{h})$ 
time that for any intensity matrix $T$ with maximum value $h$ 
finds a segmentation $\calS$ of $T$ of size at most $\frac{11}{6}
\cdot (\log_4 h +1) \cdot OPT + O(1) \cdot (\log_4 h+1),$ 
where $OPT$ is the size of a minimal segmentation of $T$.
\end{theorem}
\begin{proof}
Split $T$ into 0/1/2/3-matrices $P_{\ell}$, for $\ell=0, ...,
\log_4{h}$, 
such that $T=\sum_{\ell=0}^{\log_4{h}}4^\ell P_{\ell}$. By
Lemma~\ref{lemma_b_4}, every row of $P_{\ell}$ can be segmented using 
at most 
$\rho/2+O(1)$ 1-segments, $\rho/4$ 2-segments, and
$\rho/6$ 3-segments. Therefore, the total number of
segments required for each $P_\ell$ using {\sc GreedyRowPacking} is at
most $\rho/2 + \rho/4 + \rho/6 + O(1)$. The total number of segments
required for $T$ is then at most $(\rho/2 + \rho/4 + \rho/6 +
O(1))\cdot{}(\log_{4}h+1)$.  By Observation~\ref{observation:OPT},  $OPT\geq{}\rho/2$. Therefore, 
the size of the segmentation is at most $(\frac{11}{6} OPT +
O(1))\cdot{}(\log_4{h}+1)$ which proves the result. 
\end{proof}

Note that $\frac{11}{6}\log_4(h) < \frac{3}{2}\log_3(h) < \log_2(h)$,
so for sufficiently large 
$OPT$ and $h$, the new algorithm provides the best approximation
guarantee and is better by a factor of $\frac{12}{11}$. 
From a theoretical perspective, Theorem~\ref{theorem_b_4} is valuable because it
guarantees that solving $P_{\ell}$ matrices (either with the algorithm
implicit in Lemma~\ref{lemma_b_4} or optimally using
the results of~\cite{kalinowski:complexity}) yields an approximation guarantee. 
From an empirical perspective, preliminary experimental results 
indicated that using base $b=4$ is no better than using base $b=3$ 
in practice, and we did not pursue this approach further in our
experiments (see Section~\ref{section:empirical}).

\subsection{Even higher bases?}
\label{se:higher_base}

In theory, our approach could be taken further, using bases $b=5,6,\dots$.
There are two obstacles to doing so.  First, how to find a good segmentation
of a matrix with entries in $0,\dots,b-1$?  One can find the optimal
segmentation in time $O(mn^{2b-4})$ \cite{kalinowski:complexity}, but this
quickly becomes computationally infeasible.  Are there faster algorithms?

Secondly, would using an optimal segmentation give a good approximation?
This is not immediately clear, and in fact, the following example shows
that the approximation factor is not much better than $2\log_b(h)$.

\begin{theorem}
Consider any approximation algorithm that obtains a segmentation 
of $T$ by decomposing $T$ into $1+\log_b{h}$ matrices $P_\ell$ and 
then combining segmentations of each $P_\ell$. Any such 
algorithm can yield an approximation factor no better than 
$\frac{(2b-2)}{b}\cdot\log_b{h} + \frac{1}{b}$ in the worst case,
even for a single-row problem.
\end{theorem}
\begin{proof}
Define for $\ell=0,\dots,k-1$ matrix $P_\ell$ to be
$$(1~~2~~3~~4~~\cdots~~(b-1)~~0~~(b-1)~~\cdots{}~~3~~2~~1),$$
and set matrix $P_k$ to be
$$(0~~0~~0~~0~~\cdots~~0~~1~~0~~\cdots{}~~0~~0~~0).$$
Finally, set $T=\sum_{\ell=0}^k b^\ell P_\ell$.

Clearly $P_\ell$, for $\ell<k$ requires at least $2(b-1)$ segments
in any segmentation, and $P_k$ requires one segment, so any segmentation
of $T$ obtained with this approach has $2(b-1)k+1$ segments.
On the other hand, matrix $T$ can be segmented with just $b$ 
segments: For $i=1,\dots,b-1$, the $i$th segment has value $(011\cdots 1)_b$ 
in base $b$ and extends from column $i$ to column $2b-i$, and the $b$th segment
contains a single $1$ in the column $b$ and is otherwise 0.   Hence, a solution
obtained with this approach will have an 
approximation factor of at least $\frac{(2b-2)}{b}\cdot\log_b{h} + 
\frac{1}{b}$.
\end{proof}

If higher bases are to be used, then one way to prove an approximation
factor would be to generalize
Lemmas~\ref{lem:markers} and \ref{lemma_b_4}.  Here, we offer
the following:

\begin{conjecture}
\label{conj}
For any matrix $P_\ell$ with entries in $0,1,\dots,b-1$,  
there exists a segmentation of row $i$ of $P_\ell$ that
uses at most $\frac{1}{2v} \rho_\ell^i + O(1)$ segments of value $v$,
for $v=1,\dots,b-1$.
\end{conjecture} 

Notice that Lemmas~\ref{lem:markers} and \ref{lemma_b_4} prove this
conjecture for $b=3,4$.  If the conjecture were true, this could be used
to obtain a segmentation of $T$ of size
$(H_{b-1} OPT + O(1))(\log_b(h)+1)$, 
where $H_{b-1}=1+\frac{1}{2}+\dots+\frac{1}{b-1}$
is the {\em harmonic number}.  Since $H_{b-1}\approx \ln(b-1)$,
this means that the approximation factor is $\approx \ln(h)$ after
ignoring some lower-order terms.  

While we are not able to prove the conjecture, we can at least show 
that nothing better is possible.

\begin{lemma}
There exists a matrix $P$ with entries in $0,1,\dots,b-1$ such that
any segmentation of $P$ uses at least $H_{b-1} \cdot {\rho}/{2}$ segments.
\end{lemma}
\begin{proof}
Let $P$ be the matrix
$$
\left(
\begin{array}{cccccccc}
1 & 0 & 1 & 0 & \dots & 1 & 0 & 1 \\
2 & 0 & 2 & 0 & \dots & 2 & 0 & 2 \\
3 & 0 & 3 & 0 & \dots & 3 & 0 & 3 \\
\vdots & \vdots & \vdots & \vdots & \dots & \vdots & \vdots & \vdots \\
b-1 & 0 & b-1 & 0 & \dots & b-1 & 0 & b-1 \\
\end{array}
\right),
$$
where the number of non-zeros in each row, which is the same as $\rho/2$,
can be chosen arbitrarily.
Assume $P$ has been segmented using $n_v$ segments of value $v$.    

Consider
the $i$th row of $P$, and count not only the markers, but also the amount
by which the values at each marker change.  Thus, let $\mu_i$ be the
sum of the changes between consecutive values in row $i$; 
then $\mu_i=i\cdot \rho$.  (Similarly as for markers, changes at the leftmost
and rightmost end of the matrix are included.)
Each segment of value $v$ in row $i$ can only account for up to $2v$
change between consecutive values (namely, at its two ends).  Also notice
that necessarily $v\leq i$ since all values in row $i$ are at most $i$.
So we must have 
$$\sum_{v=1}^i 2v\cdot n_v \geq \mu_i = i \cdot \rho.$$
How small can $n_1+\dots+n_{b-1}$ be, subject to this constraint (as well
as the obvious $n_i\geq 0$ for all $i$)?  This is
a linear program, and using duality theory (see e.g. \cite{Chv83}), 
one can easily see that 
the optimal primal solution is $n_v^* = \frac{1}{v} \cdot \rho/2$.   
(The optimal dual solution
assigns $\frac{1}{i(i+1)}$ to row $i<b-1$ and $\frac{1}{b-1}$ to the last row.)
The optimal primal (and dual) solution has value $H_{b-1} \cdot \rho/2$.
While $n_v^*$ need not be integral in general, 
this nevertheless shows that any segmentation cannot be smaller than the
value of the optimal primal solution.
So any segmentation of $P_\ell, \ell<k$ requires at least
$H_{b-1}\cdot \rho/2$ segments.
\end{proof}

Note that the above matrix can in fact be segmentated using 
$\frac{1}{v}\cdot \rho/2$ segments of value $v$ if $\rho/2$ is a 
multiple of $2\cdot (b-1)!$.  What remains to do to show Conjecture~\ref{conj}
is to show that this matrix is the worst case that could happen.

We suspect that this (or a similar) matrix could also be used to devise
a target-matrix where no approximation better than $\approx ln(h)$
is possible with the split-by-base-$b$-approach, but have not been able 
to find one.

\section{Approximation by modifying row-segmentations}
\label{logD}

Our previous approximation algorithm can be summarized as follows:
split the intensity matrix by digits, split each resulting matrix
into rows, segment each row and then put the segments together.
The second approximation algorithm by Luan et al.
\cite{luan:approximation} uses another approach that is in some
sense reverse: split the intensity matrix into rows, segment each 
row, split each resulting segment into multiple segments by digits, 
and then put the segments together. The quality of this second 
approximation depends on two factors: the approximation guarantee 
and the largest value used by a segment in any of the 
row-segmentations. Without formally stating it in these terms, 
Luan et al. proved the following result:

\begin{lemma}(Luan~{\it et al.}~\cite{luan:approximation})
\label{lem:alpha}
Assume that for any single-row problem we can find an $\alpha$-approximate
solution where all segments have value at most $M$.  Then we can compute
in polynomial time an $\alpha(\log M+1)$-approximate segmentation of $T$.
\end{lemma}

Luan et al.~used this property by showing that any single-row problem has a
2-approximate solution where any segment has value at most $D$, 
where the {\em row-difference} $D$ is the maximum difference 
between consecutive elements in a row, or the maximum of the first 
and last entries in the row, whichever is larger. We can slightly 
improve on this with two observations. First, {\em any} segmentation 
can be converted into a segmentation with values at most $D$, 
without adding any new segments. Secondly, values $\alpha<2$ can 
be found in existing results.
\begin{lemma}
\label{lem:transform}
Let $\calS$ be any segmentation of a single-row intensity matrix $T$
with row-difference $D$.
Then there exists a segmentation $\calS'$ with $|\calS'|\leq |\calS|$ for which all
segments have value at most $D$.
\end{lemma}
\begin{proof}
Modify $\calS$ such that no 
two segments meet, i.e., if some segment ends at index $i$, then 
no segment starts at $i+1$. 
This can always be done ithout increasing th number of segments, 
see e.g.~\cite{baatar:decomposition}.
Any segment $S$ must then have value $v\leq 
D$, for if $S$ ends at $i$, then $T[i+1]=T[i]-v$ since no segment 
starts at $i+1$. 
\end{proof}

\vspace{-0.07cm}\noindent{}Theorem~\ref{thm:approxD} follows immediately from 
Lemma~\ref{lem:alpha} and Lemma~\ref{lem:transform}, using $M=D$: 
\vspace{-0.06cm}
\begin{theorem}\label{thm:approxD}
There exists a polynomial-time algorithm that, for any intensity 
matrix $T$ with maximum row-difference $D$, finds a segmentation 
$\calS$ of $T$ size at most $\alpha\cdot{}(\log D+1)OPT$. Here 
$\alpha \leq \frac{24}{13}\approx{}1.846$ in the general case 
by~\cite{B}.
\end{theorem}

If the running time for obtaining an $\alpha$-approximation for 
the single row problem is $t_{\alpha}$, then this algorithm runs 
in $O(t_\alpha\cdot{}m\cdot{}\log{h})$; the $\alpha \leq 
\frac{24}{13}$ algorithm can be implemented in $O(h\cdot{}n^2)$ 
time. For the general case, this approximation result improves 
upon the $2 \cdot (\log D+1)$ approximation result for the 
full-matrix problem in~\cite{luan:approximation}. In particular, 
for $\alpha=\frac{24}{13}$, if $D\leq{}(h^{13}/8)^{1/16}$, then to 
the best of our knowledge, this is the tightest approximation to 
the segmentation problem with no restriction on the intensity 
matrix values.

\vspace{-5pt}
\section{Experimental Results}\label{section:empirical} 

To examine the impact of our algorithms in practice, we implemented our new approximation 
algorithms as well as those of~\cite{luan:approximation}. In 
particular, our experiments use the following algorithms:

\begin{enumerate}

\item{\textsc{Alg}$_{b=2}$:} The $(\log_2{h}+1)$ approximation 
algorithm of~\cite{luan:approximation}.

\item{\textsc{Alg}$_{b=3}$}: The $\frac{3}{2} \cdot 
(\log_3{h}+1)$ approximation algorithm of Section~\ref{section:logh}.

\item{\textsc{Alg}$_{\alpha=2}$:} The $2(\log{D}+1)$ approximation 
algorithm of~\cite{luan:approximation}.

\item{\textsc{Alg}$_{\alpha=24/13}$:} The $\frac{24}{13} \cdot 
(\log{D}+1)$ approximation algorithm of Section~\ref{logD}, which 
utilizes our implementations of algorithms 
from~\cite{B,bar_noy:unified}.

\item{\textsc{OPT}:} The optimal solution obtained via a recent 
state-of-the-art exact algorithm of~\cite{brand:sum} which 
improves the running time over the related work 
in~\cite{baatar:minimum}.
\end{enumerate}

\noindent{}All approximation algorithms were implemented in Java 
while an implementation of \textsc{OPT} was provided as a binary executable by the author of~\cite{brand:sum}.\\

\noindent{\bf{}Scope of Our Experiments:} We restrict our 
investigation to algorithms with approximation guarantees. Aside 
from their practical performance, approximation algorithms play an 
important role by providing an efficient method for checking the 
quality of solutions provided by heuristics. While heuristics may 
perform well in practice, their lack of a performance guarantee 
means that low-quality solutions cannot be ruled out. On the other 
hand, as demonstrated by previous 
works~\cite{baatar:minimum,brand:sum} and by our experimental work 
here, computing the optimum is computationally intensive and can 
require a significant amount of time; moreover, such exact 
approaches are only possible with intensity matrices of limited 
size and $h$ values. Therefore, at the very least, approximation 
algorithms allow one to quickly verify that a heuristic is not 
producing a poor result; moreover, the approximate solution may 
indeed provide a satisfactory solution. While a comprehensive 
comparison involving the large body of literature on heuristic 
approaches would be of interest, such an undertaking is outside 
the scope of this current work.

\subsection{Data Sets}

We use the following test data:
\begin{itemize}
\item{\it Data Set I:} a real-world data set comprised of $70$ 
clinical intensity matrices obtained from the Department of 
Radiation Oncology at the University of California at the San 
Francisco School of Medicine. The levels are specified in terms of 
percentages in increments of $20$\% of some maximum value. We 
extract the common factor of $20$ to obtain values in 
$\{1,2,3,4,5\}$.

\item{\it Data Set II:} a real-world data set containing a 
prostate case, a brain case and a head-neck case obtained from the 
Department of Radiation Oncology at the University of Maryland 
School of Medicine. This data set consists of $22$ clinical 
intensity matrices with fractional values specified absolutely; 
the floor of these values are used for our experiments.

\item{\it Data Set III:} a synthetic data set of $30$ intensity 
matrices. Each matrix is obtained as follows: compute the sum of 
the probability density functions of seven bivariate Gaussians 
generated from two independent standard univariate Gaussian 
distributions where the amplitude $A$ and the centers of the 
distributions are sampled uniformly at random. The distributions 
are discretized by adding as the value in the $m\times n$-grid the 
integer part of the corresponding function value. The choice of 
seven Gaussians and the range of the amplitude (we chose 1-25) was 
made to ensure some peaks and valleys in the intensity matrix, 
while keeping the matrices reasonably small for the purposes of 
computing an optimal solution.
\end{itemize}

The utility of Data Set III is that it allows for testing on 
intensity matrices where $D$ values are relatively small compared 
to $h$. Such data allows us to address our third line of 
investigation by examining the effect of small $D$ values on the 
performance of our approximation algorithms. Moreover, testing on 
matrices with small $D$ values is pertinent assuming improvements 
in treatment technology. Higher precision MLCs can allow for more 
fine-grained intensity matrices and current technologies exist for 
supporting MLCs with up to 60 leaf pairs. Finally, we note that 
the $h$ values used in each of our data sets are fairly small - 
{\it this is necessary in order for the exact algorithm 
of~\cite{brand:sum} to complete within a reasonable amount of 
time} as we discuss in more detail later.


\subsection{Results of Experiments}

Tables~\ref{table:DS1a}-\ref{table:DS3} below contain the results 
for each instance of our experimental evaluation. All experiments 
were conducted on a machine with a $1$ GHz Pentium CPU and $1$ GB 
of RAM. In Tables~\ref{table:DS2} \& \ref{table:DS3}, the running 
times for computing the optimum are also included since these were 
significant.\vspace{-0pt}

\begin{table}[h]  
\begin{center}
\scriptsize
\hspace{-0pt}\begin{tabular}{|c|c|c|c|c|c|c|c|c|c|}
  \hline
  Instance & m & n & $h$ & $D$ & \textsc{OPT} &   \textsc{Alg}$_{b=2}$ & \textsc{Alg}$_{b=3}$ & \textsc{Alg}$_{\alpha=2}$ & \textsc{Alg}$_{\alpha=24/13}$\\
  \hline
 1  & 20 & 19   & 5 & 5 & 7 & 10  &  \underline{8}  & 12  & 12                                                    \\                                                                                                 
 2  & 19 & 18   & 5 & 5 & 8 & 11  &  \underline{9} & 11  & 11                                                     \\                   
 3  & 19 & 14   & 5 & 5 & 9 & 11  &  \underline{10}  & 15  & 15                                                   \\                      
 4  & 19 & 14   & 5 & 5 & 8 & \underline{10}  &  \underline{10}  & 13  & 15                                                   \\                     
 5  & 19 & 16   & 5 & 5 & 8 & 12  &  \underline{9}  & 14  & 13                                                    \\                       
 6  & 20 & 16   & 5 & 5 & 8 & 11  &  \underline{9}  & 12  & 12                                                    \\                      
 7  & 20 & 16   & 5 & 5 & 9 & 12  &  \underline{9}  & 14  & 15                                                    \\                      
 8  & 20 & 16   & 5 & 5 & 8 & 12  &  \underline{10}  & 13  & 13                                                   \\                      
 9  & 20 & 11   & 5 & 5 & 7 &  \underline{8}  &  \underline{8}  & 12  & 12                                        \\                      
 10 & 27 & 21   & 5 & 5 & 10 & \underline{13}  & 14  & \underline{13}  & 14                                       \\                        
 11 & 27 & 20   & 5 & 5 & 10 & 12  & 13  & \underline{11}  & \underline{11}                                       \\                      
 12 & 26 & 18   & 5 & 5 & 8 & \underline{9}  & 10   & 12  & 12                                                    \\                      
 13 & 26 & 15   & 5 & 5 & 7 & \underline{9}  &  \underline{9}   & 10  & 10                                        \\                      
 14 & 26 & 18   & 5 & 5 & 8 & \underline{11}  & 12   & 12   & 14                                                  \\                      
 15 & 26 & 17   & 5 & 5 & 8 & 11  & 11   & \underline{10}   & \underline{10}                                      \\                      
 16 & 26 & 13   & 5 & 5 & 7 & 10  & \underline{9}    & 10  & 10                                                   \\                      
 17 & 26 & 18   & 5 & 5 & 8 & \underline{11}  & \underline{11}   & \underline{11}      & \underline{11}           \\                      
 18 & 27 & 20   & 5 & 5 & 8 & 11  & \underline{10}   & \underline{10}      & \underline{10}                       \\                      
 19 & 21 & 19   & 5 & 5 & 11 & 15  & \underline{12}   & 13      & 13                                              \\                      
 20 & 21 & 17   & 5 & 5 & 7 & \underline{9}   & 10   & 12      & 12                                               \\                       
 21 & 21 & 15   & 5 & 5 & 8 & 11  & \underline{8}    & 11      & 11                                               \\                      
 22 & 20 & 18   & 5 & 5 & 9 & 12  & \underline{9}    & 14      & 14                                               \\                      
 23 & 21 & 18   & 5 & 5 & 9 & 11  & \underline{10}   & 12      & 12                                               \\                      
 24 & 21 & 15   & 5 & 5 & 6 & 8   &  \underline{7}   & 9       &  9                                               \\                      
 25 & 21 & 17   & 5 & 5 & 9 & 12  & \underline{9}    & 15      & 14                                               \\                      
 26 & 21 & 19   & 5 & 5 & 9 & 13  & \underline{10}   & 14      & 12                                               \\                      
 27 & 21 & 21   & 5 & 5 & 11 & 14  & 14   & \underline{13}      & \underline{13}                                  \\                      
 28 & 21 & 19   & 5 & 5 & 10 & 14  & \underline{13}   & \underline{13}      & \underline{13}                      \\                        
 29 & 22 & 16   & 5 & 5 & 8 & 11  & \underline{9}    & 11      & 11                                               \\                      
 30 & 21 & 11   & 5 & 5 & 5 & \underline{6}   & 7    & 7       &  7                                               \\                      
 31 & 20 & 20   & 5 & 5 & 10 & 14  & \underline{13}   & 14      & 14                                              \\                      
 32 & 20 & 19   & 5 & 5 & 9 & \underline{11}  & \underline{11}   & 12      & 13                                   \\                      
 33 & 22 & 15   & 5 & 5 & 8 & 11  & \underline{10}   & \underline{10}      & \underline{10}                       \\                      
 34 & 21 & 20   & 5 & 5 & 10 & 13  & \underline{12}   & 14      & 14                                              \\                      
 35 & 21 & 16   & 5 & 5 & 8 & \underline{9}  & \underline{9}    & 10      & 10                                    \\                      
 \hline                                                                                                                                               
\end{tabular}                                                                                                                                                                                                                                                                                                                                                      
\caption{The experimental instances 1-35 of Data Set I with the 
best result provided by the approximation algorithms 
underscored.}\label{table:DS1a}
\end{center}
\end{table}

\begin{table}[b]  
\begin{center}
\scriptsize
\hspace{-0pt}\begin{tabular}{|c|c|c|c|c|c|c|c|c|c|}
  \hline
  Instance & m & n & $h$ & $D$ & \textsc{OPT} &   \textsc{Alg}$_{b=2}$ & \textsc{Alg}$_{b=3}$ & \textsc{Alg}$_{\alpha=2}$ & \textsc{Alg}$_{\alpha=24/13}$ \\
  \hline
                36  & 21 & 14    & 5 & 5 & 8 & \underline{11}  & \underline{11} & 12 &  12      \\                                                                                                                      
                37  & 25 & 18    & 5 & 5 & 7 & \underline{10}  & \underline{10} & 11 &  \underline{10}    \\                              
                38  & 25 & 21    & 5 & 5 & 11 & 14  & \underline{13} & 14 &  \underline{13}       \\                                         
                39  & 25 & 18    & 5 & 5 & 8 & 11  & \underline{10} & 13 & 12       \\                                                       
                40  & 26 & 19    & 5 & 5 & 11 & \underline{12}  & 14 & 20 & 14      \\                                                       
                41  & 26 & 21    & 5 & 5 & 13 & 16  & \underline{15} & 19 & 17      \\                                                       
                42  & 26 & 18    & 5 & 5 & 9 & \underline{11}  & \underline{11} & 12 & 12      \\                                            
                43  & 25 & 18    & 5 & 5 & 8 & 10  & 10 & 11 & \underline{9}      \\                                                         
                44  & 25 & 17    & 5 & 5 & 8 & 11  & \underline{10} & 12 & 12      \\                                                        
                45  & 25 & 21    & 5 & 5 & 11 & 15  & \underline{12} & 15 & 15      \\                                                       
                46 & 7 & 7       & 5 & 5 & 5 &  7  & \underline{6}  & 7 & 7      \\                                                          
                47 & 7 & 8       & 5 & 5 & 4 &  6  & \underline{4}  & 7 & 7     \\                                                           
                48 & 8 & 9       & 5 & 5 & 5 &  8  & \underline{7}  & \underline{7} & \underline{7}     \\                                   
                49 & 8 & 8       & 5 & 5 & 5 &  7  & \underline{6}  & 7 & 7     \\                                                           
                50 & 8 & 9       & 5 & 5 & 5 &  7  & \underline{6}  & 7 & \underline{6}     \\                                               
                51 & 8 & 9       & 5 & 5 & 6 &  9  & \underline{7}  & 11 & 11     \\                                                         
                52 & 8 & 9       & 5 & 5 & 5 &  8  & \underline{5}  & 6 & 6     \\                                                           
                53 & 8 & 7       & 5 & 5 & 5 &  7  & \underline{5}  & 7 & 7     \\                                                           
                54 & 8 & 9       & 5 & 5 & 6 &  8  & \underline{7}  & 8 & 8     \\                                                           
                55 & 21 & 17     & 5 & 5 & 8 &  \underline{10} & \underline{10} & \underline{10} & \underline{10}  \\                         
                56 & 20 & 19     & 5 & 5 & 7 &  9  & \underline{8} & 9 & 9     \\                                                            
                57 & 19 & 14     & 5 & 5 & 5 &  7  & 8  & \underline{6} & \underline{6}       \\                                             
                58 & 20 & 18     & 5 & 5 & 7 &  \underline{7}  & 8 & 9 &  9     \\                                                           
                59 & 20 & 17     & 5 & 5 & 6 &  \underline{7}  & \underline{7} & 8 &  8      \\                                              
                60 & 19 & 15     & 5 & 5 & 3 &  5  & 6 & \underline{4} & \underline{4}      \\                                               
                61 & 20 & 18     & 5 & 5 & 8 &  \underline{9}  & 10 & 10 & 10      \\                                                        
                62 & 21 & 18     & 5 & 5 & 8 & \underline{10}  & \underline{10} & 12 & 12     \\                                             
                63 & 21 & 20     & 5 & 5 & 8 & \underline{10}  & \underline{10} & \underline{10} & \underline{10}   \\                         
                64 & 23 & 19     & 5 & 5 & 11 & 15  & \underline{12} & 16 & 16     \\                                                        
                65 & 23 & 16     & 5 & 5 & 6 & 10  & \underline{8} & \underline{8} & \underline{8}     \\                                    
                66 & 23 & 12     & 5 & 5 & 4 & \underline{6}   & \underline{6} & 7 & 7       \\                                              
                67 & 23 & 18     & 5 & 5 & 8 & 12  & \underline{10} & 13 & 11      \\                                                        
                68 & 23 & 17     & 5 & 5 & 8 &  11 & \underline{9} & 11 &  11    \\                                                          
                69 & 22 & 14     & 5 & 5 & 5 &  \underline{7}  & \underline{7} & 8 & \underline{7}     \\                                    
                70 & 22 & 16     & 5 & 5 & 7 &  \underline{8}  & 9 & 9 & 9     \\                                                            
 \hline                                                                                                                                               
\end{tabular}                                                                                                                                                                                                                                                                                                                                                      
\caption{The experimental instances 36-70 of Data Set I with the 
best result provided by the approximation algorithms 
underscored.}\label{table:DS1b}
\end{center}
\end{table}

\begin{table}[h]            
\begin{center}
\scriptsize
\vspace{-0pt}\begin{tabular}{|c|c|c|c|c|c|c|c|c|c|c|}
  \hline
  Instance & m & n  & $h$ & $D$ & \textsc{OPT} &  \textsc{Alg}$_{b=2}$ & \textsc{Alg}$_{b=3}$ & \textsc{Alg}$_{\alpha=2}$ & \textsc{Alg}$_{\alpha=24/13}$\\
  \hline
  1  & 15 & 16  & 10 & 8  & 8 (0.12)  & 18 & 15 & \underline{12} & \underline{12}      \\
  2  & 15 & 16  & 10 & 8  & 11 (0.12) & 16 & \underline{15} & \underline{15} & \underline{15}      \\
  3  & 15 & 15  & 10 & 9  & 8 (0.07)  & 15 & 16 & \underline{10} & \underline{10}      \\
  4  & 16 & 13  & 10 & 9  & 7 (0.02)  & 14 & \underline{8}  & 10 & 10      \\
  5  & 16 & 16  & 10 & 9  & 9 (0.18)  & \underline{14} & \underline{14} & \underline{14} & \underline{14}      \\
  6  & 16 & 16  & 10 & 8  & 10 (0.08) & 21 & \underline{13} & 17 & 15      \\
  7  & 15 & 13  & 10 & 10 & 5 (0.01)  & \underline{8}  & 9  & 10 &  9      \\
  8  & 23 & 27  & 10 & 9  & 14 (3.61) & 24 & \underline{21} & 25 & 25      \\  
  9  & 24 & 24  & 10 & 7  & 14 (0.32) & 21 & 18 & \underline{17} & 19      \\
 10  & 23 & 32  & 10 & 10 & 16 (1.26) & 24 & 23 & 23 & \underline{20}      \\
 11  & 23 & 24  & 10 & 8  & 14 (2.95) & 22 & 20 & \underline{19} & \underline{19}      \\
 12  & 23 & 26  & 10 & 8  & 12 (0.24) & 25 & \underline{17} & \underline{17} & 18      \\ 
 13  & 23 & 33  & 10 & 7  & 16 (2.32) & 23 & 19 & 19 & \underline{18}      \\ 
 14  & 23 & 36  & 10 & 10 & 17 (4.89) & 27 & 24 & 22 & \underline{20}      \\ 
 15  & 20 & 23  & 10 & 9  & 9 (0.12) & 14 & 14 & \underline{13} & 14      \\ 
 16  & 20 & 19  & 9  & 8  & 10 (0.02) & 14 & 16 & \underline{12} & 13      \\ 
 17  & 20 & 22  & 10 & 10 & 10 (0.08) & 15 & \underline{13} & \underline{13} & \underline{13}      \\ 
 18  & 20 & 22  & 10 & 9  & 10 (0.98) & \underline{15} & 17 & 16 & \underline{15}      \\ 
 19  & 20 & 21  & 10 & 7  & 10 (0.07) & 16 & \underline{14} & 15 & \underline{14}      \\ 
 20  & 20 & 19  & 10 & 6  & 9 (0.03) & 14 & 12 & \underline{11} & 13      \\ 
 21  & 20 & 23  & 10 & 10 & 11 (3.24) & 17 & \underline{16} & 19 & 19      \\ 
 22  & 21 & 20  & 10 & 10 & 10 (0.36) & 17 & 17 & 18 & \underline{15}      \\                                          
\hline\end{tabular} \caption{The experimental instances using Data 
Set II with the best result provided by the approximation 
algorithms underscored. The running time in CPU seconds (rounded 
to the nearest integer) for \textsc{OPT} is provided in 
parentheses.}\label{table:DS2}
\end{center}
\end{table}

\begin{table}[h]
\begin{center}
\scriptsize
\begin{tabular}{|c|c|c|c|c|c|c|c|c|c|c|}
  \hline
  Instance & m & n   & $h$ & $D$ & \textsc{OPT} &  \textsc{Alg}$_{b=2}$ & \textsc{Alg}$_{b=3}$ & \textsc{Alg}$_{\alpha=2}$ & \textsc{Alg}$_{\alpha=24/13}$\\
  \hline
  1  & 57 & 64  & 23 & 2 & 26 (21485) & 50 & 44 & 30 & \underline{29}      \\
  2  & 54 & 58  & 25 & 2 & 26 (141) & 49 & 46 & 32 & \underline{30}      \\
  3  & 57 & 58  & 24 & 2 & 23 (5) & 42 & 38 & 28 & \underline{26}      \\
  4  & 61 & 57  & 22 & 2 & 23 (17) & 42 & 42 & \underline{25} & \underline{25}      \\
  5  & 56 & 57  & 24 & 2 & 22 (1037) & 41 & 37 & \underline{25} & \underline{25}      \\
  6  & 59 & 51  & 20 & 2 & 22 (6) & 40 & 39 & \underline{23} & \underline{23}      \\
  7  & 50 & 67  & 24 & 2 & 29 (9260) & 56 & 49 & \underline{34} & \underline{34}      \\
  8  & 69 & 62  & 25 & 2 & 24 (692) & 47 & 44 & \underline{30} & \underline{30}      \\  
  9  & 62 & 64  & 18 & 2 & 19 (2) & 36 & 34 & \underline{20} & 21      \\
 10  & 59 & 59  & 23 & 2 & 28 (120822) & 54 & 49 & \underline{32} & \underline{32}      \\
 11  & 51 & 51  & 23 & 2 & 21 (15) & 40 & 37 & 25 & \underline{22}      \\
 12  & 59 & 60  & 23 & 2 & 25 (8) & 47 & 46 & 28 & \underline{27}      \\ 
 13  & 49 & 50  & 23 & 2 & 20 (25) & 38 & 35 & 26 & \underline{25}      \\ 
 14  & 59 & 45  & 23 & 2 & 19 (104) & 34 & 33 & \underline{22} & \underline{22}      \\ 
 15  & 46 & 53  & 18 & 2 & 22 (2) & 42 & 40 & 27 & \underline{23}      \\ 
 16  & 53 & 63  & 21 & 2 & 22 (11) & 45 & 40 & \underline{24} & \underline{24}      \\ 
 17  & 49 & 66  & 24 & 2 & 24 (848) & 45 & 41 & \underline{29} & \underline{29}      \\ 
 18  & 64 & 64  & 25 & 2 & 24 (6) & 44 & 43 & 33 & \underline{31}      \\ 
 19  & 53 & 53  & 25 & 2 & 22 (121) & 41 & 40 & 27 & \underline{25}      \\ 
 20  & 51 & 57  & 25 & 2 & 23 (564) & 45 & 42 & 28 & \underline{24}      \\ 
 21  & 50 & 46  & 24 & 2 & 19 (3) & 35 & 33 & 26 & \underline{22}      \\ 
 22  & 61 & 58  & 24 & 2 & 25 (5060) & 48 & 44 & \underline{26} & \underline{26}      \\
 23  & 57 & 62  & 19 & 2 & 22 (3) & 43 & 38 & 26 & \underline{22}      \\
 24  & 58 & 65  & 21 & 2 & 26 (53) & 51 & 44 & \underline{27} & 29      \\
 25  & 59 & 45  & 24 & 2 & 21 (4) & 38 & 35 & \underline{26} & \underline{26}      \\ 
 26  & 54 & 50  & 15 & 2 & 19 (1) & 34 & 33 & \underline{20} & \underline{20}      \\ 
 27  & 67 & 61  & 20 & 2 & 17 (3) & 32 & 29 & \underline{19} & \underline{19}      \\ 
 28  & 63 & 64  & 25 & 2 & 26 (506) & 50 & 46 & \underline{31} & \underline{31}      \\ 
 29  & 54 & 60  & 18 & 2 & 21 (1) & 43 & 38 & 24 & \underline{23}      \\
 30  & 63 & 58  & 24 & 2 & 23 (317) & 45 & 42 & 26 & \underline{25}      \\                                      
\hline\end{tabular}\caption{The experimental instances using Data 
Set III with the best result provided by the approximation 
algorithms underscored. The running time in CPU seconds (rounded 
to the nearest integer) for \textsc{OPT} is provided in 
parentheses.}\label{table:DS3}
\end{center}
\end{table}
 
\clearpage

\subsection{Analysis \& Discussion} 

Table~\ref{table:trial_summary} summarizes the performance of our 
approximation algorithms by enumerating the number of instances in 
which each algorithm outperformed all others (excluding 
\textsc{OPT}) with ties included.\vspace{-5pt}
\begin{table}[h]                        
\vspace*{2mm} \hspace*{\fill}
\begin{tabular}{|l|c|c|c|c|c|}
\hline
& \# Instances &  \textsc{Alg}$_{b=2}$ & \textsc{Alg}$_{b=3}$ & \textsc{Alg}$_{\alpha=2}$ &  \textsc{Alg}$_{\alpha=24/13}$  \\
\hline \hline
Data Set I & 70 & ~~~~24 ~(34.3\%) &  ~~~~{\bf{}55} ~(78.6\%)& ~~~~14 ~(20.0\%) & ~~~~18 ~(25.7\%)\\
\hline
Data Set II & 22 & ~~~~3 ~~(13.6\%) & ~~~~~9 ~~(40.9\%) &  ~~~~11 ~(50.0\%) & ~~~~{\bf{}12} ~(54.5\%) \\
\hline
Data Set III & 30 & ~~0 ~~(0.0\%) & ~~~0 ~~(0.0\%)& ~~~~16 ~(53.3\%) &  ~~~~{\bf{}28} ~(93.3\%)  \\
\hline
\end{tabular}
\hspace*{\fill} \vspace*{2mm} \caption{The number of instances 
where each of approximation algorithms achieves the smallest 
segmentation with ties included. The largest value in each row is 
bolded.} \label{table:trial_summary}\vspace{-5pt}
\end{table}

\noindent{}In testing our algorithms, we focus on three questions:
\begin{enumerate}
\item{}How do our improved algorithms compare against their older counterparts in~\cite{luan:approximation}?

\item{}How do the algorithms with an $O(\log{h})$ approximation guarantee compare to those
with an $O(\log{D})$ approximation guarantee?

\item{}How do these approximation algorithms compare against the optimum solution?
\end{enumerate}

\noindent{\bf Question 1:} With respect to our first question, 
Table~\ref{table:trial_summary} illustrates that 
\textsc{Alg}$_{b=3}$ and \textsc{Alg}$_{\alpha=24/13}$ outperform 
on a larger number of instances than the algorithms 
of~\cite{luan:approximation} in all three data sets for a total of 
$95$ out of $122$ instances (77.8\%). In particular, 
\textsc{Alg}$_{b=3}$ ties or outperforms all other approximation 
algorithms in $55$ out of the $70$ instances (78.5\%) in Data Set 
I while \textsc{Alg}$_{\alpha=24/13}$ ties or outperforms all 
other approximation algorithms in $12$ out of the $22$ instances 
(54.5\%) in Data Set II and in $28$ out of the $30$ instances 
(93.3\%) in Data Set III. We also enumerate the number of times 
one of our new algorithms outperforms an older algorithm on an 
instance-by-instance basis; this comparison is summarized in 
Table~\ref{table:question1a} along with ties (percentages along a 
row may not sum exactly to 100\% due to rounding). The results 
indicate that our new algorithms perform better than their older 
counterparts on a significant number of instances.

\begin{table}[h] \vspace*{2mm} \hspace*{\fill}
\begin{tabular}{|c|l|l|l|} \hline
  & ~\textsc{Alg}$_{b=2}$ outperforms \textsc{Alg}$_{b=3}$ & ~~\textsc{Alg}$_{b=3}$ outperforms \textsc{Alg}$_{b=2}$  & ~~~~~~~ Ties\\
\hline \hline
Data Set I & \hspace{40pt}12 ~(17.1\%) & \hspace{40pt}40 ~(57.1\%)  & \hspace{6pt}18 ~(25.7\%)\\
\hline
Data Set II & \hspace{45pt}4 ~(18.2\%) & \hspace{40pt}15 ~(68.2\%) & \hspace{10pt}3 ~(13.6)\\
\hline
Data Set III & \hspace{45pt}0 ~(0.0\%) & \hspace{40pt}29 ~(96.7\%) & \hspace{10pt}1 ~(3.3\%) \\
\hline \hline
 & ~\textsc{Alg}$_{\alpha=2}$ outperforms \textsc{Alg}$_{\alpha=\frac{24}{13}}$  & ~~\textsc{Alg}$_{\alpha=\frac{24}{13}}$ outperforms \textsc{Alg}$_{\alpha=2}$ & ~~~~~~~Ties \\
\hline \hline
Data Set I & \hspace{45pt}5 ~(7.1\%) & \hspace{40pt}12 ~(17.1\%) & \hspace{6pt}53  ~(75.7\%)\\
\hline
Data Set II & \hspace{45pt}5 ~(22.7\%) & \hspace{45pt}8 ~(36.4\%) & \hspace{10pt}9 ~(40.9\%) \\
\hline
Data Set III & \hspace{45pt}2 ~(6.7\%) & \hspace{40pt}14 ~(46.7\%) & \hspace{6pt}14 ~(46.7\%) \\
\hline
\end{tabular}
\hspace*{\fill} \vspace*{0mm} \caption{An instance-by-instance 
comparison of old vs. new $O(\log{h})$ algorithms, old vs. new 
$O(\log{D})$ algorithms.} \label{table:question1a}\vspace{-5pt}
\end{table}

\noindent{}Given these positive results, we also wish to know by 
{\it how much} we improve. We look at the number of segments 
required by an algorithm per instance and calculate the ratio of 
these two values; the average (Ave.), median (Med.), minimum 
(Min.) and maximum (Max.) ratios over all instances is reported in 
Table~\ref{table:v_values}. These values demonstrate that 
\textsc{Alg}$_{b=3}$ performs substantially better than 
\textsc{Alg}$_{b=2}$ overall judging by both the average and 
median values. In the case of \textsc{Alg}$_{\alpha=24/13}$ and 
\textsc{Alg}$_{\alpha=2}$, our gains are smaller, yet we still 
observe a small overall improvement judging by the average 
values.\\ \vspace{-5pt}

\begin{table}[h]                        
\vspace*{2mm} \hspace*{\fill}
\begin{tabular}{|c|c|c|c|}
\hline
\multicolumn{2}{|c|}{} & Ratio of \textsc{Alg}$_{b=3}$ over \textsc{Alg}$_{b=2}$ & Ratio of \textsc{Alg}$_{\alpha=\frac{24}{13}}$ over \textsc{Alg}$_{\alpha=2}$  \\
\hline \hline
\multirow{4}{*}{Data Set I} & Ave. & 0.9262   &  0.9860 \\
& Med. & 0.9161 &  1.0000 \\
& Min. & 0.6250 &  0.7000 \\
& Max. & 1.2000 &  1.1667 \\
\hline
\multirow{4}{*}{Data Set II} & Ave.  & 0.9074  & 0.9878  \\
& Med. & 0.8990 &  1.0000\\
& Min. & 0.5714 &  0.8333\\
& Max. & 1.1429 &  1.1818\\
\hline
\multirow{4}{*}{Data Set III} & Ave. & 0.9280  & 0.9650 \\
& Med. & 0.9230 &  1.0000 \\
& Min. & 0.8627 &  0.8462 \\
& Max. & 1.0000 &  1.0741  \\
\hline
\end{tabular}
\hspace*{\fill} \vspace*{2mm} \caption{Average, median, minimum 
and maximum ratios measuring the extent of our improvements.} 
\label{table:v_values}\vspace{-5pt}
\end{table}

\noindent{\bf Question 2:} Next we address our second question 
regarding the performance of the algorithms with an $O(\log{h})$ 
approximation guarantee versus those with an $O(\log{D})$ 
approximation guarantee. We restrict ourselves to a comparison of 
\textsc{Alg}$_{b=3}$ and \textsc{Alg}$_{\alpha=24/13}$ given the 
results of the previous discussion. Table~\ref{table:question2a} 
provides the results of our comparison on an instance-by-instance 
basis. As before, we also calculate the average, median, minimum 
and maximum ratios on a per-instance basis of 
\textsc{Alg}$_{\alpha=24/13}$ over \textsc{Alg}$_{b=3}$; these 
statistics are in Table~\ref{table:question2b}.\vspace{-5pt}

\begin{table}[h]                        
\vspace*{2mm} \hspace*{\fill}
\begin{tabular}{|c|c|c|c|}
\hline
 & \textsc{Alg}$_{b=3}$ outperforms \textsc{Alg}$_{\alpha=\frac{24}{13}}$ & \textsc{Alg}$_{\alpha=\frac{24}{13}}$ outperforms \textsc{Alg}$_{b=3}$  & Ties\\
\hline \hline
Data Set I & ~47 (67.1\%) & 6 (8.6\%) & ~17 (24.3\%)\\
\hline
Data Set II & ~~7 (31.8\%) & ~9 (40.9\%) &  ~~6 (27.3\%)\\
\hline
Data Set III & 0 (0.0\%) & 30 (100.0\%) & 0  (0.0\%)\\
\hline
\end{tabular}
\hspace*{\fill} \vspace*{2mm} \caption{An instance-by-instance 
comparison of \textsc{Alg}$_{b=3}$ and 
\textsc{Alg}$_{\alpha=24/13}$.} 
\label{table:question2a}\vspace{-5pt}
\end{table}

\begin{table}[h]                        
\vspace*{2mm} \hspace*{\fill}
\begin{tabular}{|c|c|c|c|c|}
\hline
 & Average & Median  & Minimum  & Maximum   \\
\hline \hline
Data Set I    & 1.1650 & 1.1111 &  0.4444 & 1.8889 \\

Data Set II   & 0.9810 & 1.000  & 0.6250  & 1.2500  \\

Data Set III  & 0.6413 & 0.6526 &  0.5714 & 0.7429 \\
\hline
\end{tabular}
\hspace*{\fill} \vspace*{2mm} \caption{Average, median, minimum 
and maximum ratios of \textsc{Alg}$_{\alpha=24/13}$ over 
\textsc{Alg}$_{b=3}$.} \label{table:question2b}\vspace{-5pt}
\end{table}

We can tentatively draw some conclusions from our analysis. We 
observe that when $h$ and $D$ are relatively equal, the 
$\frac{3}{2}\cdot{}(\log_3{h}+1)$ approximation can yield superior 
performance in practice judging by both the instance-by-instance 
comparison in Table~\ref{table:question2a} and the average and 
median values of Table~\ref{table:question2b}; this is certainly 
the case for Data Set I. However, as Data Set II illustrates, 
there are exceptions and neither algorithm is clearly superior 
here. For the case where $D$ is significantly smaller than $h$, 
all statistics suggest that the $24/13\cdot{}(\log{D}+1)$ 
approximation can yield substantially
better solutions.\\

\noindent{\bf Question 3:} We address our third question by 
examining the performance of our approximation algorithms against 
the optimum number of segments. Table~\ref{table:question3a} 
provides the average, the median, the worst, the best, and the 
{\it best}~(the smallest) theoretical approximation factor achieved 
by each algorithm over each data set. We observe that the 
theoretical values appear pessimistic as our approximation 
algorithms generally do much better. We also note that the 
theoretical approximation values for \textsc{Alg}$_{b=3}$ are 
worse than that of \textsc{Alg}$_{b=2}$ since $h$ and $OPT$ 
are not sufficiently large for our theoretical improvements to 
emerge. Relatively small $h$ values are required in order to 
compute the optimum; however, we still observe improved 
performance from \textsc{Alg}$_{b=3}$ despite the pessimistic 
approximation guarantee. Moreover, we observe that the 
approximation algorithms never exceed an approximation factor of 
$2.25$ in practice and the other statistics demonstrate that the 
approximation factor can be significantly lower. Indeed, by 
executing all four approximation algorithms, we never exceed an 
approximation factor of $1.80$ (this {\it worst} case occurs in 
Data Set II with \textsc{Alg}$_{\alpha=24/13}$) over all instances 
in all data sets. Such computations can be performed easily since 
these algorithms incur low computational overhead. By performing 
such an operation and taking the {\it best} performance on an 
instance-by-instance basis, the statistics presented in 
Table~\ref{table:question3b} can be obtained. In conclusion, the 
statistics in Tables~\ref{table:question3a} and 
\ref{table:question3b} show that these algorithms can provide very 
good approximations to the optimum.\\

\begin{table}[h]                        
\vspace*{2mm} \hspace*{\fill}
\begin{tabular}{|c|l|c|c|c|c|}
\hline
\multicolumn{2}{|c|}{} &  \textsc{Alg}$_{b=2}$ & \textsc{Alg}$_{b=3}$  & \textsc{Alg}$_{\alpha=2}$  &\textsc{Alg}$_{\alpha=24/13}$   \\
\hline \hline
\multirow{5}{*}{Data Set I}  &  Average & 1.34  & 1.23   &  1.44 & 1.41\\

  & Median  & 1.37 & 1.24  & 1.4  & 1.39 \\

  & Worst & 1.67 & 2.00  &  1.83 & 1.87 \\

  & Best & 1.00 & 1.00  & 1.10  & 1.10 \\

  & Theory & 3.32  & 3.79  & 6.64  & 6.13\\
\hline
\multirow{4}{*}{Data Set II} &  Average & 1.66  & 1.49  & 1.47 & 1.44\\

  & Median & 1.56  & 1.43  & 1.43 &  1.44 \\

  & Worst  & 2.25  & 2.00  & 2.00  &  1.80 \\

  & Best   & 1.40  & 1.14 & 1.19  &  1.12 \\

  & Theory & 4.17  & 4.65 & 7.17  & 6.62 \\
\hline
\multirow{4}{*}{Data Set III}  & Average & 1.90  & 1.76  & 1.17 & 1.13\\

  & Median  & 1.90  & 1.76  & 1.17 & 1.12\\

  & Worst  & 2.05  &  1.84  & 1.40 & 1.29 \\

  & Best  & 1.79  &  1.65  & 1.04 & 1.00 \\

  & Theory & 4.90   & 5.29   &  4.00 & 3.69 \\
\hline
\end{tabular}
\hspace*{\fill} \vspace*{2mm} \caption{Statistics on the 
approximation factors achieved by the approximation algorithms.} 
\label{table:question3a}\vspace{-5pt}
\end{table}

\begin{table}[h]                        
\vspace*{2mm} \hspace*{\fill}
\begin{tabular}{|c|c|c|c|c|}
\hline
 & Average & Median  & Worst  & Best   \\
\hline \hline
Data Set I & 1.19  & 1.18 &  1.50 & 1.00 \\

Data Set II  &  1.35 & 1.36  & 1.60  & 1.13  \\

Data Set III  &  1.12 & 1.12  &  1.29 & 1.00 \\
\hline
\end{tabular}
\hspace*{\fill} \vspace*{2mm} \caption{Statistics on the {\it 
best} approximation factor achieved by running all approximation 
algorithms on each instance of a data set and taking the best result.} 
\label{table:question3b}\vspace{-5pt}
\end{table}

\noindent{\bf Running Time:} Finally, we note {\it the running 
times of the approximation algorithms are negligible}. In 
particular, all approximation algorithms completed each instance 
within {\it at most} $0.01$ CPU seconds on Data Set I, $0.02$ CPU 
seconds on Data Set II, and $0.240$ CPU seconds on Data Set III. 
In contrast, the running time for computing an optimal solution 
can be significant. For Data Set II, the algorithm 
of~\cite{brand:sum} runs in a reasonable amount of time. However, 
recall that the values in this data set are rounded down - this 
was done to ensure that an optimal solution could be computed. 
While incorporating another decimal place of the data values 
improves the accuracy of the treatment solution, the resulting 
intensity matrices {\it simply cannot be solved optimally in any 
reasonable amount of time} due to an $h$ value that has now become 
one order of magnitude larger; this is a concern for present-day 
real-world instances. From a more forward-looking perspective, 
larger intensity matrices may become feasible as technology 
advances (MLCs with $60$ leaf pairs currently exist); however, 
increasing the dimensions of the matrix also increases the running 
time of the exact algorithm. The impact of these two factors 
begins to become apparent in Data Set III where computing an 
optimal solution for certain test cases requires substantial CPU 
time (hundreds to thousands of CPU seconds - see 
Table~\ref{table:DS3}) for moderately larger 
matrices and for $h\leq{}25$. Therefore, while exact algorithms 
like~\cite{brand:sum} are an extremely valuable approach to 
solving these problems, their utility may be limited.

\section{Conclusion}\vspace{-5pt}
We provided new approximation algorithms for the full-matrix 
segmentation problem. We first showed that the single-row 
segmentation problem is fixed-parameter tractable in the largest 
value of the intensity matrix. Using this yields provably good 
approximate segmentations for the full matrix, after suitably 
splitting either the intensity matrix or approximate segmentations 
of its rows according to some base-$b$ representation. Finally, 
our experimental results demonstrate that our theoretical 
improvements yield new algorithms that, in both the $O(\log{h})$ 
and $O(\log{D})$ cases, significantly outperform previous 
approximation algorithms in practice and can achieve reasonable 
approximations to the optimal solution, especially if executed in 
concert.

It may be of interest to explore the case of $b\geq 4$. Can 
approximation algorithms that perform better in practice be obtained?
Are further heuristic improvements  
possible, such that empirical performance in practically relevant 
cases is increased, while maintaining desirable theoretical 
approximation guarantees? Can we more exactly determine the 
threshhold where the $O(\log{h})$ approximation and $O(\log{D})$ 
approximation lead to differing performance in practice? Finally, 
a comprehensive comparison of heuristic and approximation 
algorithms is an interesting avenue of future work.\\

\noindent{\bf{}Acknowledgements:} We thank Baruch Schieber for his 
helpful correspondence with regard to~\cite{B}. We also thank 
Sebastian Brand and Natashia Boland for graciously providing the 
implementation work in~\cite{brand:sum}. Finally, we thank Samuel 
Fiorini for bringing to our attention the work done 
in~\cite{kalinowski:complexity}.

\clearpage
\end{document}